# A ROOT/IO Based Software Framework for CMS

William Tanenbaum
*FNAL, Batavia, IL 60510, USA*

The implementation of persistency in the Compact Muon Solenoid (CMS) Software Framework uses the core I/O functionality of ROOT. We will discuss the current ROOT/IO implementation, its evolution from the prior Objectivity/DB™ implementation, and the plans and ongoing work for the conversion to "POOL", provided by the LHC Computing Grid (LCG) persistency project.

## 1. INTRODUCTION

The CMS experiment [1] is one of the four approved LHC experiments. Data taking is scheduled to begin in 2007, and will last at least ten years. The CMS software and computing task [2] will be 10-1000 times larger than that of current HEP experiments. Therefore it is essential that software must be modular, flexible, and maintainable as well as providing high performance and quality.

One of the technologies utilized has been a C++ based object oriented database management system (ODBMS). Originally, the specific implementation used for object persistency was a commercial product, Objectivity/DB [3]. In 2001, it became apparent that Objectivity was not the optimal long term solution for data persistency, and that it was necessary to abandon Objectivity with a very short time scale. A decision was made to directly use ROOT/IO [4] as a component of an interim persistency implementation. In the very near future, the LHC computing grid persistency project will provide POOL [5] as an implementation for persistency.

This paper primarily covers the conversion from Objectivity/DB to ROOT/IO. Also briefly discussed is the ongoing transition to POOL.

## 2. ROOT/IO BASED COBRA

The CMS Software Framework (COBRA), formerly called the CMS Analysis and Reconstruction Framework (CARF), is described in detail elsewhere [6]. This paper deals only with the low level (technology dependent) persistency storage management aspects of COBRA. A higher level discussion of the CMS approach to persistency can be found elsewhere [7]. Suffice it to say that communication with the persistent data store is handled by COBRA rather than explicitly by user written simulation, reconstruction, or analysis packages.

### 2.1. General Design of ROOT/IO COBRA

A decision was made to minimize the dependence of COBRA on the persistency implementation details. As a result, standard STL container classes (e.g. std::vector, std::string) are used throughout rather than ROOT specific classes (e.g., TClonesArray, TString). Also, the use of the ClassDef() macro is kept to an absolute minimum. The only optional ROOT specific class that is used is TRef, the ROOT class supporting references to persistent objects.

For simplicity, it was decided at this stage to use ROOT/IO for all persistent data, including metadata. This greatly standardized and simplified the conversion. Also, for simplicity, ROOT/IO folders, trees and branches are not used.

### 2.2. Objectivity/DB to ROOT Conversion

As mentioned above, the original version of COBRA used Objectivity/DB as its implementation of persistent objects. Many Objectivity specific features (e.g. namescopes) were used extensively throughout COBRA. So pervasive was the influence of Objectivity that it was decided that it was not feasible to redesign COBRA to stop using every Objectivity specific feature in the short time available to us. Rather, where necessary, an Objectivity specific feature would be emulated with ROOT/IO. In essence, an Objectivity emulator is implemented for those Objectivity capabilities that could not easily be removed or replaced.

Below, we discuss the mappings from Objectivity/DB to ROOT/IO.

#### 2.2.1. The Federated Database

An Objectivity Federated Database (a.k.a. federation) (class ooFDObj) is a collection of user defined databases and the associated schema.

No specific ROOT/IO analogue of a federation is used. The absence of a file catalog or similar structure to tie together ROOT files is acceptable for the interim ROOT solution. The POOL file catalog will provide this capability in the future.

#### 2.2.2. Objectivity databases

A ROOT/IO file (class TFile) is used in place of each Objectivity database (class ooDBObj), with a 1-1 correspondence between them.

#### 2.2.3. Objectivity containers

A ROOT/IO directory (class TDirectory) is used in place of each Objectivity container (class ooContObj).

#### 2.2.4. Objectivity objects

A ROOT/IO named object (class TNamed) is used in place of each Objectivity persistent object (class ooObj). Although Objectivity objects are unnamed, it is necessary to name ROOT/IO objects to support persistent references. As the names need be unique only within a container, it was decided to use human-readable mnemonic names for the objects rather than machine generated unique object identifiers. This made it easier to use ROOT interactively to examine a file.

### 2.2.5. Objectivity vectors and iterators

STL vectors and iterators are used respectively in place of each Objectivity persistent vector (class ooVArray<T>) and each Objectivity iterator (class ooItr(T)).

### 2.2.6. Objectivity namescopes

Each Objectivity object or container has its own name space in which any other object or container may be given a unique name. Objectivity calls these name spaces "namescopes". Objectivity supports bidirectional access to namescopes, i.e. a name can be accessed either through the scoping object or the named object. In COBRA with ROOT/IO, persistent objects that need to support namescopes do not inherit directly from TNamed. Rather, they inherit indirectly through an intermediate class that contains an STL map and an STL multimap containing the bidirectional namescope information. In order to support namescopes for containers, each container contains a keyed "namescope" object containing the map and multimap.

### 2.2.7. Objectivity persistent references

Objectivity supports references to persistent objects, containers, or databases (class ooRef(T)). These references are used to locate a persistent object in memory or to retrieve it from the persistent store if it is not in memory. ROOT uses the TRef class to support references to persistent objects. However, a TRef can only be used to uniquely identify an object in memory. A TRef does not provide the capability to locate an object in the persistent store.

In order to provide a reference that can uniquely identify and retrieve an object in memory or in the persistent store, COBRA defines a persistent class that contains a TRef and also the names of the ROOT/IO file, container, and object in the persistent store. This is why named objects (class TNamed) are used.

### 2.2.8. Objectivity transactions

ROOT/IO does not support atomic transactions. However, COBRA with ROOT/IO mimics transactions by keeping a record of all objects to be written, and writing them out at definite user-defined intervals. In addition, a small "master" collection object is the last object to be written at each interval. If a crash or other interruption occurs prior to the writing of the master object, other written or partially written objects will be overwritten when the run is resumed. Hence the window of vulnerability is limited to the interval during the writing of the master record.

## 2.3. Scale of conversion effort

The conversion effort was done by a single developer, new to COBRA, over a five month period, two months for coding and three months for developer debugging and testing.

## 2.4. Performance

The ROOT/IO framework typically uses about half the disk space of the Objectivity based framework. The default ROOT compression level (level 1) is used. Considering that CMS data will be on the order of petabytes, this is a huge improvement.

There is an increase of a few percent in the running time of event digitization production jobs due mainly to the overhead of data compression.

## 3. POOL BASED COBRA

There are several deficiencies to the current ROOT/IO based COBRA. The most important of these is the absence of a file catalog, making large-scale production difficult. Other deficiencies include the lack of a cache manager, and the ad-hoc solution to the problem of references to persistent objects. POOL [5] solves all of these problems, as well as decoupling COBRA from any specific persistency technology. The conversion of COBRA to POOL is currently in progress.

## Acknowledgments

The author wishes to thank Vincenzo Innocente for his help in understanding COBRA, Rene Brun, Philippe Canal, Masaharu Goto, and Fons Rademakers for their prompt resolution of ROOT issues, Walter Brown for his C++ expertise, and Veronique Lefebure for her skill in finding bugs in the ROOT/IO based framework.

Work supported by the U. S. Department of Energy.